# Large Photoelasticity in Topological Antiferromagnet $Mn_3Sn$ Revealed by Coherent Acoustic Phonon

Yuchen Wang[A], Takuya Matsuda[A,B], Yuta Murotani[A], Hanyi Peng[B], Takumi Matsuo[B], Tomoya Higo[B], Satoru Nakatsuji[A,B,C,D], and Ryusuke Matsunaga[A,C]

[A] *The Institute for Solid State Physics, The University of Tokyo, Kashiwa, Chiba 277-8581, Japan*

[B] *Department of Physics, The University of Tokyo, Bunkyo-ku, Tokyo, 113-0033, Japan*

[C] *Trans-scale Quantum Science Institute, The University of Tokyo, Bunkyo-ku, Tokyo, 113-0033, Japan*

[D] *Institute for Quantum Matter and Department of Physics and Astronomy, Johns Hopkins University, Baltimore, 21218 Maryland, USA*

**Abstract:**

We investigate the role of ultrafast strain on the electronic and optical responses in topological antiferromagnet $Mn_3Sn$ thin films using near-infrared femtosecond pump-probe spectroscopy. Coherent acoustic phonons are generated and exhibit remarkably large oscillations in differential transmission exceeding 1% in amplitude. Our quantitative analysis reveals that $Mn_3Sn$ possesses an unusually large near-infrared photoelastic coefficient, several times greater than those in conventional materials, indicating a remarkable sensitivity of the electronic states to lattice distortions. This work establishes a quantitative framework for understanding and utilizing strain-induced responses in $Mn_3Sn$, paving a foundation for exploring coupled electron-phonon-magnon dynamics for ultrafast straintronics.

Antiferromagnets (AFMs) have recently attracted significant interest due to their potential for spintronics with higher integration density and ultrafast processing speed, because of the absence of stray field and the fast precessional motion of AFM spins in terahertz (THz) regime[1–4]. Particularly noteworthy is the non-collinear topological Weyl AFM $Mn_3Sn$[5]. Its crystalline structure is characterized by an ABAB stacking sequence of kagome bilayers, where Mn spins within each layer form an inverse triangular structure below $T_N$~430 K[6–9], which explicitly breaks the time-reversal symmetry (See Fig. 1(a)) despite negligibly small net magnetization. This unique spin structure gives rise to exceptionally large room-temperature responses across electrical, optical, and thermal channels, including a large anomalous Hall effect[5,10], a strong magneto-optical Kerr effect in the visible light range[11], a sizable Faraday effect in the THz regime[12], a pronounced anomalous Nernst effect[13], and tunneling magnetoresistance in all-antiferromagnetic tunnel junctions[14]. The combination of ultrafast spin dynamics and the pronounced electronic responsiveness to external stimuli at room temperature makes $Mn_3Sn$ an exceptionally promising platform for AFM-based spintronics[15]. A key next challenge is to achieve ultrafast manipulation of the AFM order. Recently, electric current-induced magnetic switching[16–20] and chiral spin rotation[21,22] via spin-orbit torque in $Mn_3Sn$-based heterostructures have been reported as a promising pathway toward high-speed control of AFM spins. While these approaches demonstrate effective control of AFM order, they rely on charge transport and may introduce Joule heating[19,23], motivating the exploration of alternative control schemes. In this context, strain provides an appealing route, as it directly couples to the magnetic order via the piezomagnetic effect. A recent study reported that switching the AFM order in bulk $Mn_3Sn$ can be achieved by applying static in-plane uniaxial strain on the order of ~0.1%[24]. In addition, substrate-induced strain in $Mn_3Sn$ thin films has also been shown to play a crucial role in current-induced full switching between AFM binary states[18]. These findings highlight the importance of strain and its considerable effect on magnetism[25–27].

Recently, the speed of straintronics has been pushed up into the picosecond regime. Lattice distortions can be induced to control the spin states via nonlinear phonon excitation by intense laser pulses[28–30] and the circular motion of ions in the lattice or substrate[31,32]. Even in metals, where excitation of infrared-active phonons is suppressed by electronic screening, femtosecond optical pulses offer an alternative route to generate ultrafast strain in the form of coherent acoustic phonon (CAP), *i.e.*, elastic wave[33–35]. CAP-induced modulation of magnetization in ferro- and ferrimagnets[36–40] has been reported. Whether such ultrafast strain-mediated control operates in $Mn_3Sn$ remains an open question. Addressing this requires a quantitative understanding of how strain couples to the optical and electronic responses in $Mn_3Sn$. Generally, ultrafast changes in magnetization can be probed by time-resolved magneto-optical Kerr effect (TR-MOKE) because the MOKE signal $\theta_K$ is proportional to the magnetization: $\theta_K = \alpha \times M$, where $M$ is the magnetization and $\alpha$ is a proportional coefficient that depends on the complex refractive index. However, it should be noted that a TR-MOKE signal $\Delta\theta_K$ is not necessarily proportional to a change in magnetization $\Delta M$ because $\Delta\theta_K = \Delta\alpha \times M + \alpha \times \Delta M$. This relation indicates that, when the refractive index is significantly modulated by strain ($\Delta\alpha \neq 0$), a non-zero TR-MOKE signal ($\Delta\theta_K \neq 0$) can arise even if the magnetization itself remains unchanged ($\Delta M = 0$). Indeed, a recent time-resolved terahertz Faraday rotation study on $Mn_3Sn$ reported that the ultrafast sub-ps change in Faraday rotation induced by femtosecond laser pulses is not explained by the change in the magnetic order[41]. Therefore, for careful interpretation of the ultrafast strain-driven effects, it is critical to evaluate photoelasticity, *i.e.,* a strain-induced modulation of the optical refractive index. In addition, recent reports[42] on kagome staircase oxides have revealed large photostrictive responses. These effects arise from the pronounced structural anisotropy and buckled kagome geometry, which make the lattice exceptionally responsive to light-induced perturbations. Such observations suggest that kagome frameworks may support enhanced strain-optical coupling. This motivates exploring how strain modulates the ultrafast optical response in metallic kagome systems, including $Mn_3Sn$. The study of CAP and quantitative evaluation of photoelastic properties would provide valuable information for optoacoustic properties in this novel material.

In this work, we perform near-infrared (NIR) pump-probe spectroscopy to investigate the CAP dynamics and strain-induced optical response in $Mn_3Sn$ thin films. We observe oscillations in the differential transmittance ($\Delta T/T$) with an amplitude as large as 1% in a 20 nm-thick film, and a frequency dependent on film thickness. The result is successfully reproduced by our analysis modeling the CAP generated by photoinduced heating in the films. Remarkably, the real part of the photoelastic coefficient in $Mn_3Sn$ reaches approximately 24, several times larger than that of typical materials. Our quantitative analysis of the optical response in $Mn_3Sn$ reveals its remarkable sensitivity to strain, together with a robust linearity even under intense photoirradiation. These results establish a quantitative connection between ultrafast strain and the electronic response, providing a foundation for strain-mediated ultrafast control of antiferromagnetic order.

As depicted in Fig. 1(b), we used 15-40 nm-thick polycrystalline $Mn_3Sn$ films grown on 500 μm-thick quartz ($SiO_2$) substrates[43] (see Supplemental Material S1 for details). Pump pulses at 1030 nm (160 fs, 3 kHz) from Yb:KGW laser (PHAROS, Light Conversion) were incident on the sample at a small angle from the surface normal, with a spot diameter of 0.45 mm. Probe pulses of the same wavelength and duration, with a smaller spot diameter 0.16 mm and fluence of 50 μJ $cm^{-2}$, were employed to measure the pump-induced changes in transmittance as a function of time delay (see Supplemental Material S2). All measurements were performed at room temperature unless otherwise noted.

Figure 2(a) shows representative results of transient differential transmission signals $\Delta T/T$ of a 20 nm-thick $Mn_3Sn$ thin film. A pronounced oscillation in $\Delta T/T$ emerged at approximately 10 ps after photoexcitation. The non-oscillatory background was fitted with a bi-exponential function including finite rise time (dashed line), and the residual oscillatory part was subsequently extracted. The oscillation $\Delta T_{\rm osc}/T$ is well fitted by

$$\frac{\Delta T_{\rm osc}}{T} = Ae^{-\frac{t}{\tau_d}}\cos(2\pi f_{\rm osc}t + \varphi), \tag{1}$$

where $A$, $\tau_d$, $f_{\rm osc}$, and $\varphi$ represent the amplitude, decay time, frequency, and initial phase of the oscillation, respectively. We varied the probe polarization (data not shown) and found no dependence on the polarization angle. The same approach was applied to samples of different thicknesses. Fig. 2(b) shows the oscillation amplitude as a function of pump fluence. Remarkably, the amplitude exhibits a robust linear dependence on the pump fluence over the entire experimental range, up to approximately 3 mJ $cm^{-2}$. In the thinner sample, the oscillation amplitude reaches and even exceeds 1%. Such a large and linear relationship is uncommon compared to typical materials, which will be discussed later.

Figure 2(c) presents the $\Delta T/T$ dynamics for films with various thicknesses. As the film thickness decreases, the oscillation frequency increases. Using Eq. (1), we extracted the corresponding oscillation frequency $f_{\rm osc}$ and damping rate $1/\tau_d$, as plotted by solid triangles and circles, respectively, in Fig. 2(d). Both $f_{\rm osc}$ and $1/\tau_d$ scale approximately inversely with the film thickness. This behavior can be understood within a simple one-dimensional (1D) longitudinal standing wave model inside the film. The corresponding frequency of the $N$th mode is given by $f_{\rm sw}^{(N)} = Nv/2d$, where $N$ is a positive integer, $v$ is the speed of sound in the sample, and $d$ is the film thickness. Using the density $\rho$=7440 kg/$m^3$, Young's modulus $E\sim139$ GPa, and Poisson's ratio $\nu_p$=0.129 for $Mn_3Sn$[44], we estimated $v = \sqrt{E(1-\nu_p)/\rho(1-\nu_p-2{\nu_p}^2)}$=4.4 nm/ps. The solid curve in Fig. 2(d) shows the calculated $f_{\rm sw}^{(1)}$ for the fundamental ($N = 1$) mode, which is in good quantitative agreement with the experiment. Considering the acoustic impedance mismatch between the sample and substrate, the vibration can be described as a quasi-two-free-end standing wave, while part of the acoustic energy is transmitted into the substrate. This behavior is consistent with the CAP generation via impulsive photoexcitation with finite decay.

To quantitatively describe the CAP dynamics, we develop the following model based on thermoelastic generation[34], as shown by Fig. 3(a). A Gaussian-shaped 1030-nm pulse irradiates the $Mn_3Sn$ film and acts as the transient heat source. Since the spot sizes of both the pump and probe beams are much larger than the film thickness, a 1D model is sufficient to simulate the process. In the following, $z$ denotes the direction normal to the sample surface. We first consider the heat equation,

$$\left(\frac{\partial}{\partial t}-\frac{k}{\rho c_p}\frac{\partial^2}{\partial z^2}\right)\Delta T_K(z,t)=s(z,t), \tag{2}$$

where $c_p$=431 J/(kg·K)[44], $\Delta T_K(z,t)$, and $s(z,t)$ denote the specific heat capacity, lattice temperature rise, and the laser-induced heat source, respectively. In practice, the absorbed pulse energy initially couples to the electronic system. Our calculations based on the two-temperature model indicate that the electron and lattice temperatures equilibrate within several picoseconds, consistent with recent experimental observations[41]. Since we focus on longer timescales, we approximate the temperature rise by considering only the lattice temperature (see Supplemental Material S3 for details). Once the heat source is quantified, $\Delta T_K(z,t)$ can be obtained. We then consider the elastic wave equation,

$$\left(\frac{\partial^2}{\partial t^2}-v^2\frac{\partial^2}{\partial z^2}-\frac{\eta}{\rho}\frac{\partial^3}{\partial z^2\partial t}\right)u(z,t)=-\frac{\beta K}{\rho}\frac{\partial\Delta T_K(z,t)}{\partial z}, \tag{3}$$

where $\eta$=$10^{-3}$ Pa·s (see details in Supplementary Material S4), $\beta$=$3.8\times10^{-5}$ $K^{-1}$ [45], $K$=$62\times10^9$ Pa[44], and $u(z,t)$ denote the effective viscosity coefficient, linear expansion coefficient, bulk modulus, and the displacement induced by the temperature rise, respectively. From this, we obtain the values of the displacement $u(z,t)$ at a certain position $z$, as well as the longitudinal strain $\eta_{33}(z,t)=\partial u(z,t)/\partial z$, stress $\sigma_{33}(z,t)$, etc. Additionally, we account for the boundary conditions that include continuity of heat flux and stress at the film-substrate interface. The latter governs the damping of CAP, i.e., energy transfer into the substrate, due to acoustic impedance mismatch. At the film surface in contact with air, a free boundary condition is assumed, i.e., $\sigma_{33}(0,t)$=0.

Figure 3(b) presents the calculated relative thickness change $\Delta d/d$ in the 25-nm sample, *i.e.*, $[u(25,t)-u(0,t)]/d$. The calculated dynamics show a zigzag-shaped oscillation with a thermal expansion background, closely resembling the differential transmittance in Fig. 2(a). The waveform indicates a superposition of odd higher-order modes ($N=3,5,7,\ldots$) allowed by the boundary conditions, a common feature in laser-induced CAP[46]. In contrast, the experimentally observed oscillation evolves toward a sinusoidal shape from the second cycle, indicating a faster damping of the higher-order modes. This behavior deviates from the case of perfect adhesion between film and substrate, where all modes decay at the same rate, and instead suggests the presence of frequency-dependent damping channels in the real scenario[47,48]. The thickness-dependent oscillation frequency and damping rate of $\Delta d_{\mathrm{osc}}/d$ are extracted and plotted in Fig. 2(d) as open triangles and circles, respectively, to compare with the experiment. The calculated damping rate shows reasonable agreement with experimental results and scales with oscillation frequency. Furthermore, the relation $\tau_d\propto d$ indicates that energy transmission to the substrate is the dominant contribution to the damping[47,48]. Overall, the model quantitatively reproduces both the oscillation frequency and the damping rate of $\Delta T_{\mathrm{osc}}/T$, providing a consistent description of CAP generation and dissipation in $Mn_3Sn$ thin films under sub-picosecond perturbations.

Figure 3(c) shows two-dimensional maps of the calculated temperature rise (upper panel) and strain (lower panel) as functions of position $z$ and delay time $t$. Figure 3(d) shows spatial cross-sections of the strain at selected delay time, illustrating its time evolution. The maximum strain, corresponding to the peak indicated by the arrow in Fig. 3(b), reaches up to 0.3% at the highest fluence. Previous research showed that ~0.1% static strain is sufficient to flip the spin orientation[24]. This comparison highly suggests that the optically generated strain is large enough to influence the magnetic state. In the present experiment, however, the sample temperature is calculated to be elevated from 300 to 468 K, exceeding $T_N$. Lowering the base temperature[49] or attaching a thermoelastic transducer[36–40] may provide promising pathways toward realizing ultrafast strain-induced control of AFM spins, which will be explored in future work.

In this work, we focus on the remarkably large differential transmittance, reaching up to 1% in Fig. 2(a). From Fig. 3(b), the relative thickness change at the oscillation maximum is estimated to be on the order of $\Delta d/d \sim 10^{-3}$, comparable to values reported in previous CAP studies. This is expected, as the elastic parameters[44] of $Mn_3Sn$ used in Eq. (3) are not significantly different from those of typical materials. In contrast, the oscillation amplitude of the differential transmittance is remarkably large. This comparison indicates that the large oscillation observed in Fig. 2 cannot be attributed to thickness modulation alone but instead originates from an enhanced strain-induced optical response. This response is quantified by photoelasticity, characterized by the complex coefficient $\partial\tilde{n}_1/\partial\eta_{33}$, which describes strain-induced change in complex refractive index $\tilde{n}_1 = n_1 + i\kappa_1$. Based on the theory of the photoelasticity[33,50,51], the strain-induced transient differential transmission (see derivation in Supplemental Material S5) could be written as

$$\frac{\Delta T}{T} = 2\mathrm{Re}\left\{ ik_0 \frac{\partial\tilde{n}_1}{\partial\eta_{33}} \left[ \frac{r_1}{1 + r_0 r_1 e^{2i\delta_0}} \int_0^d \eta_{33}(z,t) e^{2ik_1(d-z)} dz - \frac{r_0}{1 + r_0 r_1 e^{2i\delta_0}} \int_0^d \eta_{33}(z,t) e^{2ik_1 z} dz + \left( \frac{1 - r_0 r_1 e^{2i\delta_0}}{1 + r_0 r_1 e^{2i\delta_0}} \right) \int_0^d \eta_{33}(z,t) dz \right] \right\}. \quad (4)$$

where $k_0$ is the probe wave vector in vacuum, $r_0$ and $r_1$ are the amplitude reflection coefficients of air/$Mn_3Sn$ and $Mn_3Sn/SiO_2$ interfaces, respectively, and $\delta_0 = k_0\tilde{n}_1 d$ is the complex optical phase in the absence of strain. Note that the integral of strain, in the third term on the right-hand side, represents the thickness change, which macroscopically reflects the strain distribution. To solve Eq. (4) using experimentally obtained $\Delta T_{\mathrm{osc}}/T$, two unknown parameters are involved: the real- and imaginary-parts of photoelastic coefficient, $\partial n_1/\partial\eta_{33}$ and $\partial\kappa_1/\partial\eta_{33}$. We perform a global fit to the data from films of different thicknesses, assuming a common, thickness-independent photoelastic coefficient (see details in Supplemental Material S6). The strain distribution for each thickness is used as a known input to Eq. (4), with the photoelastic coefficient as the only unknown complex global fitting parameter. Figure 4(a) shows the global fitting results for 20, 25, 30, and 40 nm films, demonstrating good agreement with the experimental results. The average value is obtained from 1000 fitting runs and compared with previously reported values for other materials[51–53], as plotted in Fig. 4(b). The photoelastic coefficient in $Mn_3Sn$ is determined to be $\partial\tilde{n}_1/\partial\eta_{33} \sim 23.9(\pm1.2) - i8.4(\pm1.4)$, which is several times larger than that of conventional metals.

To examine the relation between the large photoelastic response and its topologic feature, we investigated the temperature dependence of CAP. Figure 4(c) plots the oscillation amplitude as a function of the base temperature for the 15-nm film. At around 270 K, where a spin reorientation phase transition to a helical phase occurs[54], the amplitude starts to decrease, and drops to ~65% at 200 K. Notably, at temperatures close to or even higher than $T_N$, the oscillation amplitude remains large, suggesting that the large photoelasticity is not specific to the inverse-triangular AFM phase. This implies that the large photoelastic response may be a more general feature of kagome materials, in which the electronic responses can be strongly influenced by lattice structure with flat bands and resultant strong electronic correlation[55–59]. Even in the paramagnetic phase, calculation shows that a flat band close to Fermi level exists in $Mn_3Sn$[57]. However, in helical phase, the band structure is dramatically modified[60]. The large photoelasticity in $Mn_3Sn$ is a manifestation of the sensitivity of its electronic states to lattice distortion, which facilitates strain-based modulation of its electronic and optical responses.

In summary, we observe large differential transmittance oscillations exceeding 1% in photoexcited $Mn_3Sn$ thin films, exhibiting a robust linear dependence on pump fluence. Quantitative analysis based on heat-driven CAP reproduces the experimental results and reveals the remarkably large near-infrared photoelastic coefficient of $Mn_3Sn$ compared to typical materials. The large strain-induced optical response provides a sensitive probe for tracking the ultrafast post-pump dynamics. These results suggest that ultrafast strain provides an efficient pathway for modulating and investigating the electronic states in kagome magnets. Our work establishes a quantitative framework for ultrafast strain manipulation in $Mn_3Sn$ films, paving the way for future studies on coupled electron-phonon-magnon dynamics.

## SUPPLEMENTARY MATERIALS

Supplementary materials include sample preparation, experimental setup, theoretical model and calculation of the CAP, derivation of the relationship between $\Delta T/T$ and CAP, and more details.

## AUTHOR DECLARATIONS

### Conflict of Interest

The authors have no conflicts to disclose.

### Author Contributions

This work was supported by JST Mirai Program (Grant No. JPMJMI20A1), JSPS KAKENHI (Grant No. JP24K16990), and JST SPRING (Grant No. JPMJSP2108). T.M. and R.M. conceived this project. H.P., T.M., T.H., and S.N. fabricated the samples. Y.W. and T.M. constructed the optical setup and performed the experiments. Y.M. and Y.W. conducted theoretical analysis. Y.W. and R.M. wrote the manuscript with substantial feedback from T.M., Y.M., T.H., and all the coauthors.

## DATA AVAILIBILITY

The data that support the findings of this study are available from the corresponding author upon reasonable request.

## Figures and legends

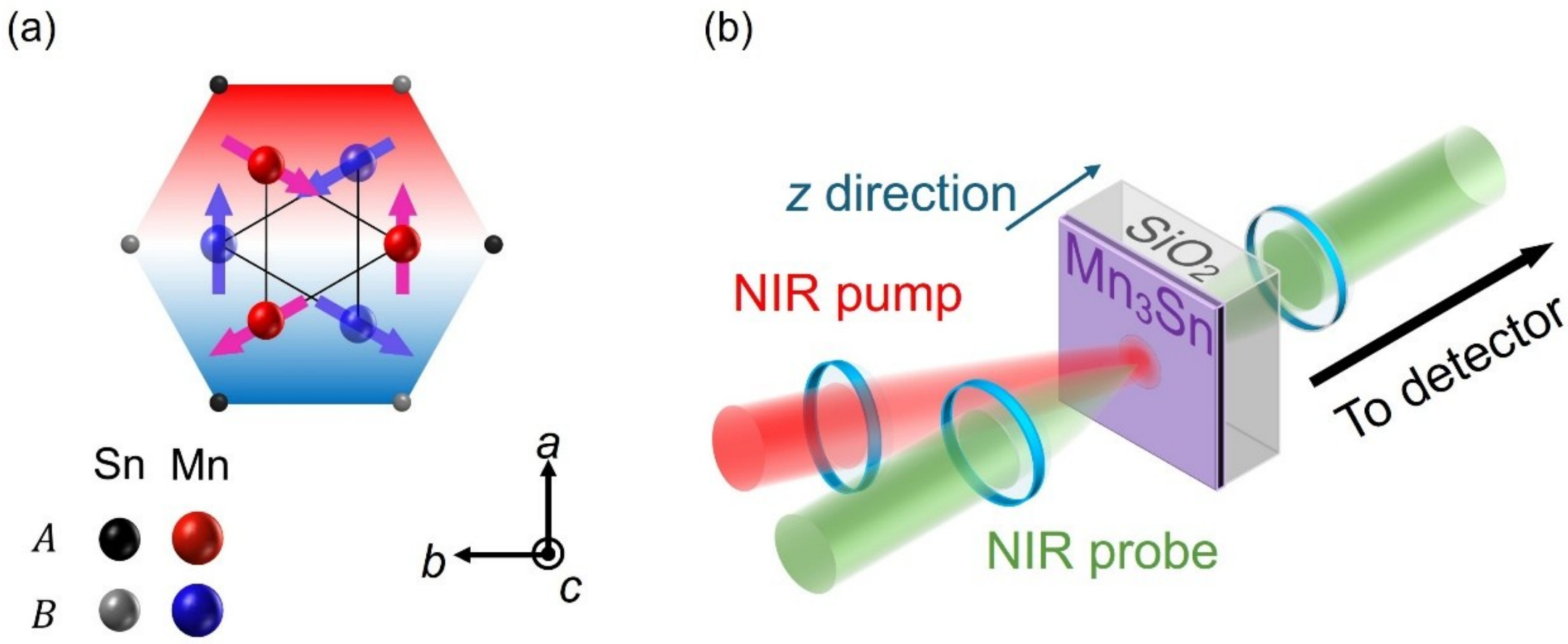

**Fig. 1.** (a) Crystal and magnetic structure of the non-collinear antiferromagnet $Mn_3Sn$. The magenta and blue arrows represent the spin of Mn atoms in A and B layer, respectively. (b) Schematic diagrams of the sample/substrate structure and the NIR pump-probe spectroscopy configuration.

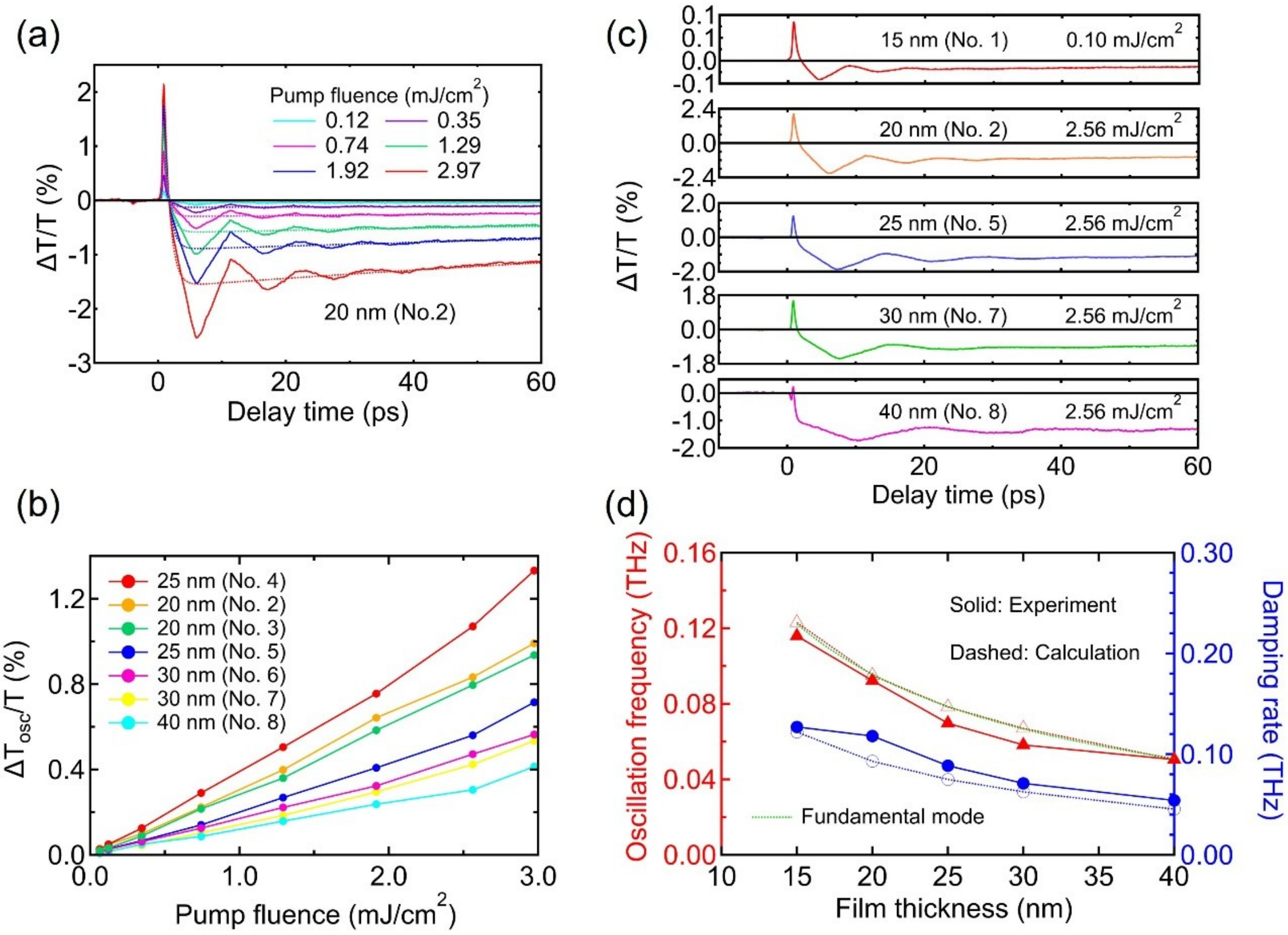

**Fig. 2.** (a) Transient differential transmission $\Delta T/T$ as a function of delay time in 20 nm-thick $Mn_3Sn$ thin film at different pump fluences. (b) Dependence of peak-to-zero oscillation amplitude on pump fluence. (c) $\Delta T/T$ dynamics in 15-40 nm films. (d) Dependence of oscillation frequency and damping rate on film thickness. The solid lines with filled markers and dashed lines with open markers show results of the experimental observations of $\Delta T_{osc}/T$, and theoretical calculations of $\Delta d_{osc}/d$, respectively. The green dashed line shows the fundamental frequency of the standing wave.

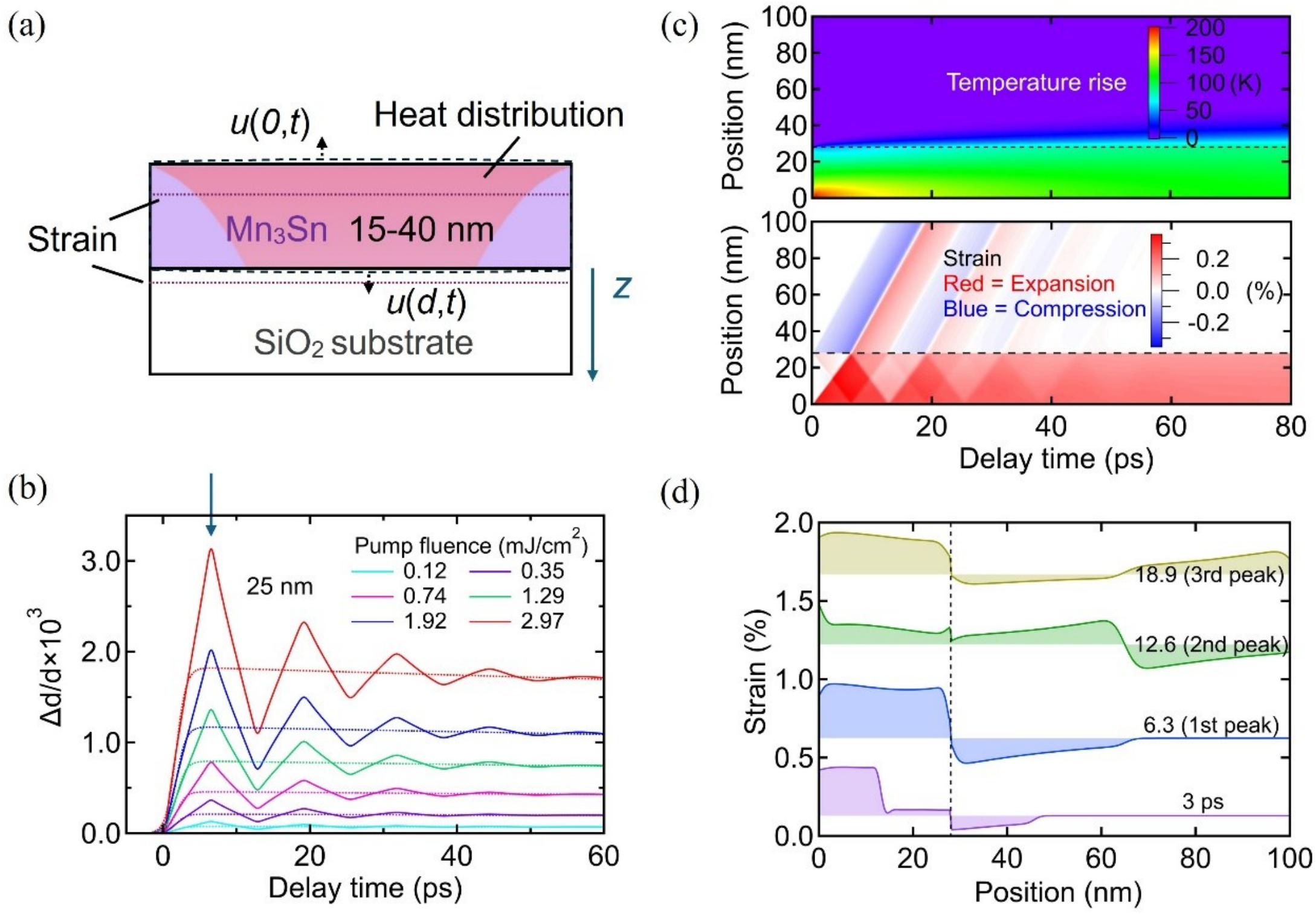

**Fig. 3.** (a) Two-dimensional view of the photoexcited film structure. The strain field is generated in both the $Mn_3Sn$ and $SiO_2$ substrate. For simplicity, a $\delta$-function strain profile is shown. (b) Calculation of the thickness change in 25 nm sample. The arrow points at the first peak. (c) Time-dependent evolution of the temperature (upper) and strain (lower) in the 25 nm sample and substrate. The dashed black lines represent the interface. (d) Spatial cross-sections of the strain evolution at various delays.

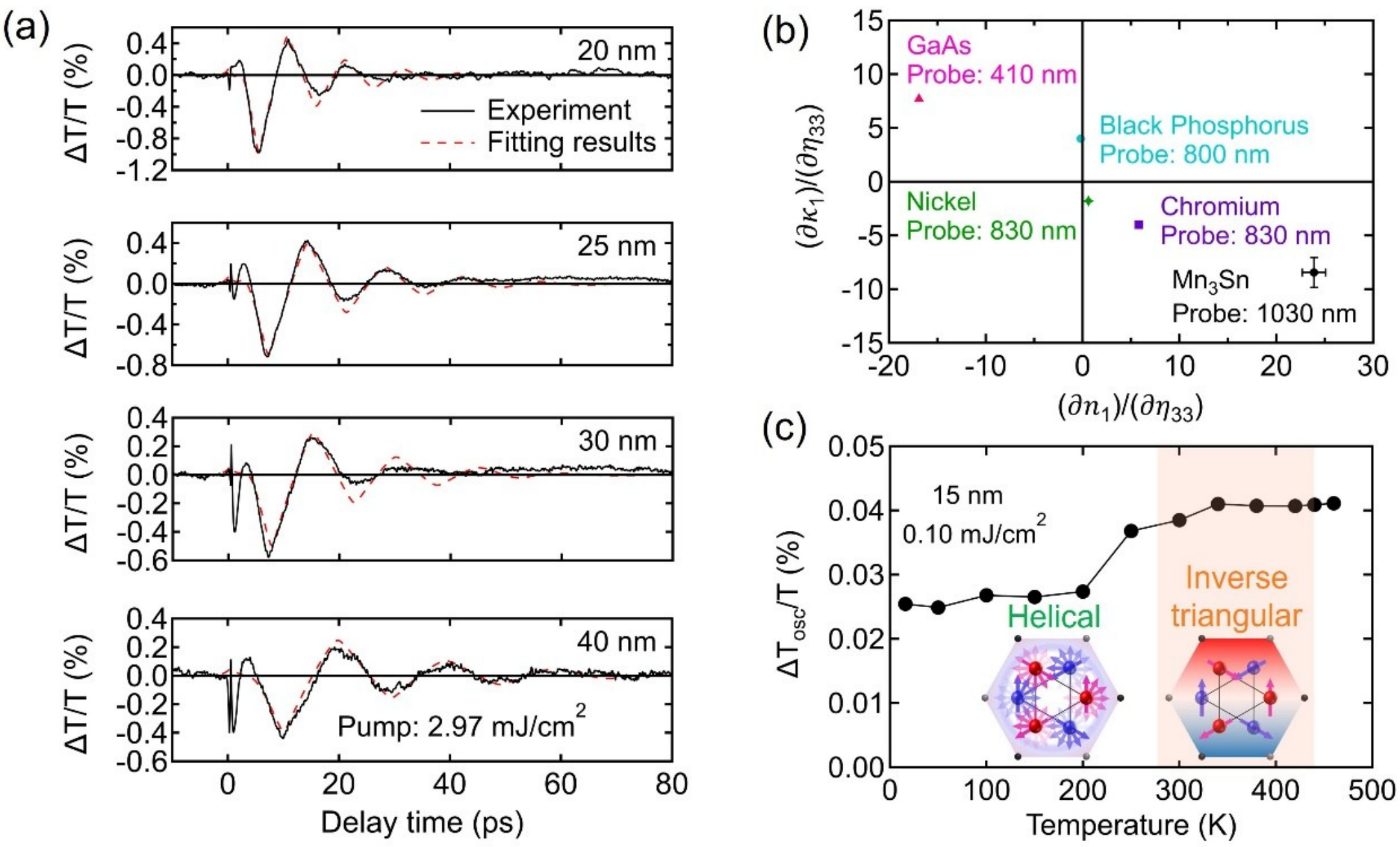

**Fig. 4.** (a) Global fitting results of 20 nm (No. 2), 25 nm (No. 5), 30 nm (No. 7), and 40 nm (No. 8). The black solid lines are the experimental results, and red dashed lines correspond to global fitting results obtained by Eq. (4). (b) Complex photoelastic coefficient of $Mn_3Sn$ thin films and comparison with other materials. (c) Temperature dependence of oscillation amplitude in 15 nm film. A clear decrease at ~250 K is observed. Inset panels show the top views of the magnetic order at each phase.

# Large Photoelasticity in Topological Antiferromagnet $Mn_3Sn$ Revealed by Coherent Acoustic Phonon: Supplementary Materials

Yuchen Wang[A], Takuya Matsuda[A,B], Yuta Murotani[A], Hanyi Peng[B], Takumi Matsuo[B], Tomoya Higo[B], Satoru Nakatsuji[A,B,C,D], and Ryusuke Matsunaga[A,C]

[A] *The Institute for Solid State Physics, The University of Tokyo, Kashiwa, Chiba 277-8581, Japan*
[B] *Department of Physics, The University of Tokyo, Bunkyo-ku, Tokyo, 113-0033, Japan*
[C] *Trans-scale Quantum Science Institute, The University of Tokyo, Bunkyo-ku, Tokyo, 113-0033, Japan*
[D] *Institute for Quantum Matter and Department of Physics and Astronomy, Johns Hopkins University, Baltimore, 21218 Maryland, USA*

## S1. Sample fabrication

$Mn_3Sn$ polycrystalline thin films (15-40 nm) were deposited on 500 μm-thick quartz ($SiO_2$) substrates by DC magnetron sputtering from a $Mn_{2.7}Sn$ alloy target in a chamber with a base pressure of $< 5\times10^{-7}$ Pa. An $AlO_x$ passivation layer (3 nm) was subsequently deposited by RF magnetron sputtering. Both the $Mn_3Sn$ and $AlO_x$ layers were prepared at room temperature, followed by annealing at 500 °C for 30 minutes. The sputtering power was 60 W for $Mn_3Sn$, and 100 W for $AlO_x$. The composition of the $Mn_3Sn$ layers was determined as $Mn_{3.1}Sn_{0.9}$ for sample No. 1 (15 nm), No. 2 (20 nm), No. 5 (25 nm), No. 7 (30 nm) and $Mn_{2.98}Sn_{1.02}$ for sample No. 3 (20 nm), No. 4 (25 nm), No. 6 (30 nm), No. 8 (40 nm) by scanning electron microscopy-energy dispersive X-ray spectrometry (SEM-EDX).

## S2. Experimental setup

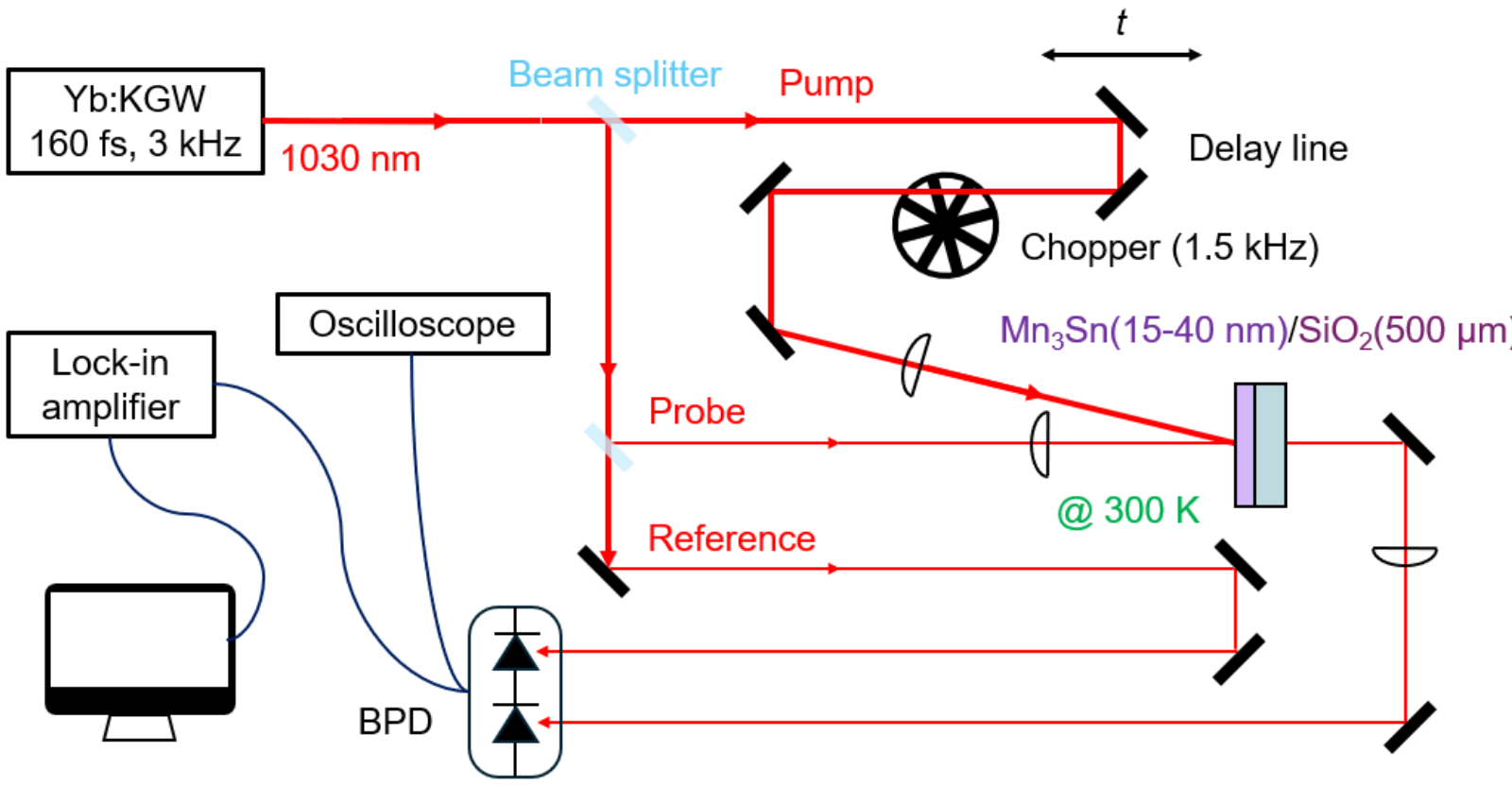

Fig. S1. Experimental setup for pump-probe differential spectroscopy. BPD: balanced photodiode detector.

### S3. Two-temperature model

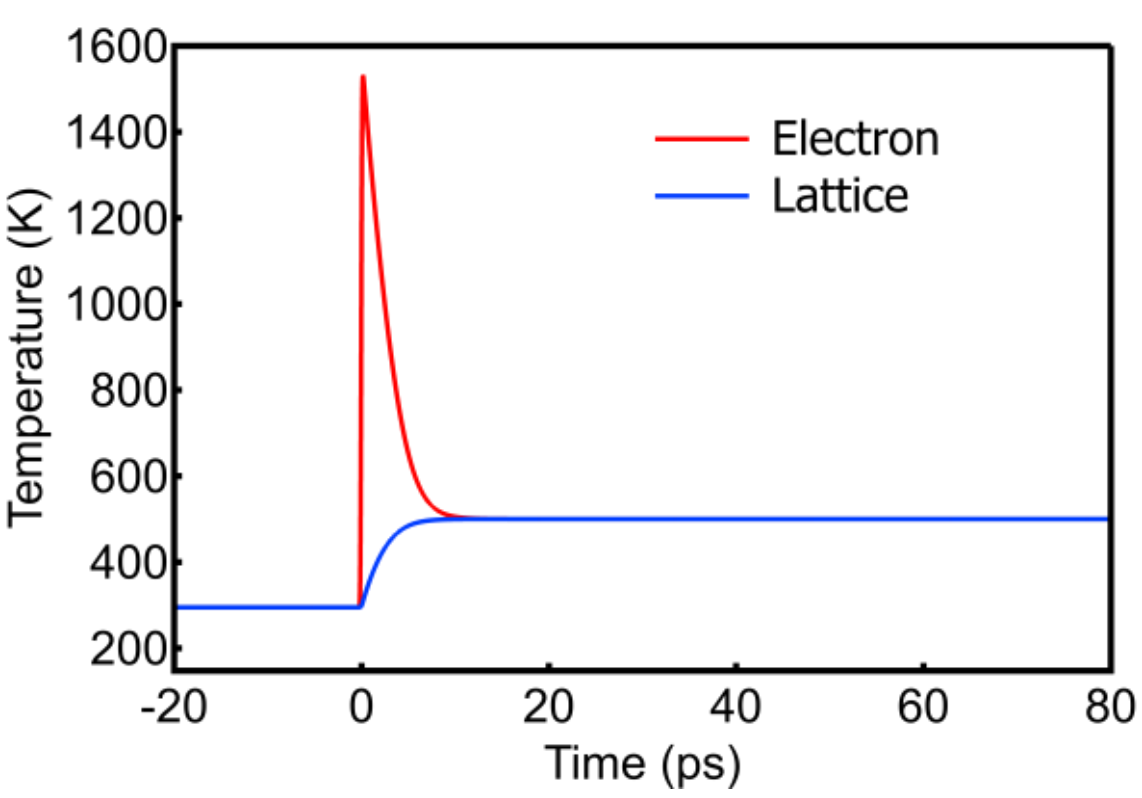

Fig. S2. Calculation of electron and lattice temperature.

Figure S2 shows the calculation of electron temperature $T_e$ and lattice temperature $T_l$ based on the two-temperature model:

$$C_e \frac{\partial T_e}{\partial t} = s(t) - g(T_e - T_l), \tag{S1}$$

$$C_l \frac{\partial T_l}{\partial t} = g(T_e - T_l), \tag{S2}$$

where $C_e = 596$ J m$^{-3}$ K$^{-1}$, $C_l = 3.3\times10^6$ J m$^{-3}$ K$^{-1}$, and $g = 1.8\times10^5$ J ps$^{-1}$ m$^{-3}$ K$^{-1}$ [S1, S2] represents the electron heat capacity, lattice heat capacity, and electron-phonon coupling coefficient, respectively. Here the initial temperature is set to 295 K and the pump fluence is 3 mJ cm$^{-2}$. The calculation shows that the $T_e$ and $T_l$ reach thermal equilibrium within approximately 10 ps. Since the generation of CAP is driven by lattice expansion, and we focus on oscillatory behavior, only $T_l$ is considered in the CAP calculation. The sub-picosecond peak observed in the $\Delta T/T$ in the main text is attributed to the ultrafast electronic response, which follows the rapid rise of $T_e$, as illustrated by the red curve in Fig. S2.

### S4. Evaluation of viscosity and its influence on CAP oscillation amplitude

In the main text, we use an effective viscosity of $\eta$=10$^{-3}$ Pa·s for $Mn_3Sn$ films. Here, $\eta$ represents an effective damping parameter describing acoustic attenuation. We refer to the values of mechanical quality factor $Q_m$ (at 2 GHz) of some metallic materials such as Pt (5400), W (3100), Au (1700), Mo (1100), Ti (690), reported in ref. [S3]. Generally, in metals, $Q_m \propto f^{-1}$, below 10 GHz, while the decrease becomes much slower at higher frequencies. Hence, we conservatively estimate $Q_m$ of $Mn_3Sn$ to be ~100 to 1000 at 10 GHz, and ~50 to 500 at 100 GHz. Using the Voigt relation $\eta = \rho v^2/\omega Q_m$ [S3, S4], we obtain $\eta$~5×10$^{-4}$ to 5×10$^{-3}$ Pa·s at 100 GHz. Accordingly, we adopt $\eta$=10$^{-3}$ Pa·s. As shown in Fig. S3, varying the viscosity from 10$^{-4}$ to 10$^{-3}$ Pa·s results in only minor changes in $\Delta d/d$ oscillation amplitude, indicating that the extracted photoelastic coefficient and main conclusions are not sensitive to the exact value of $\eta$ within this range.

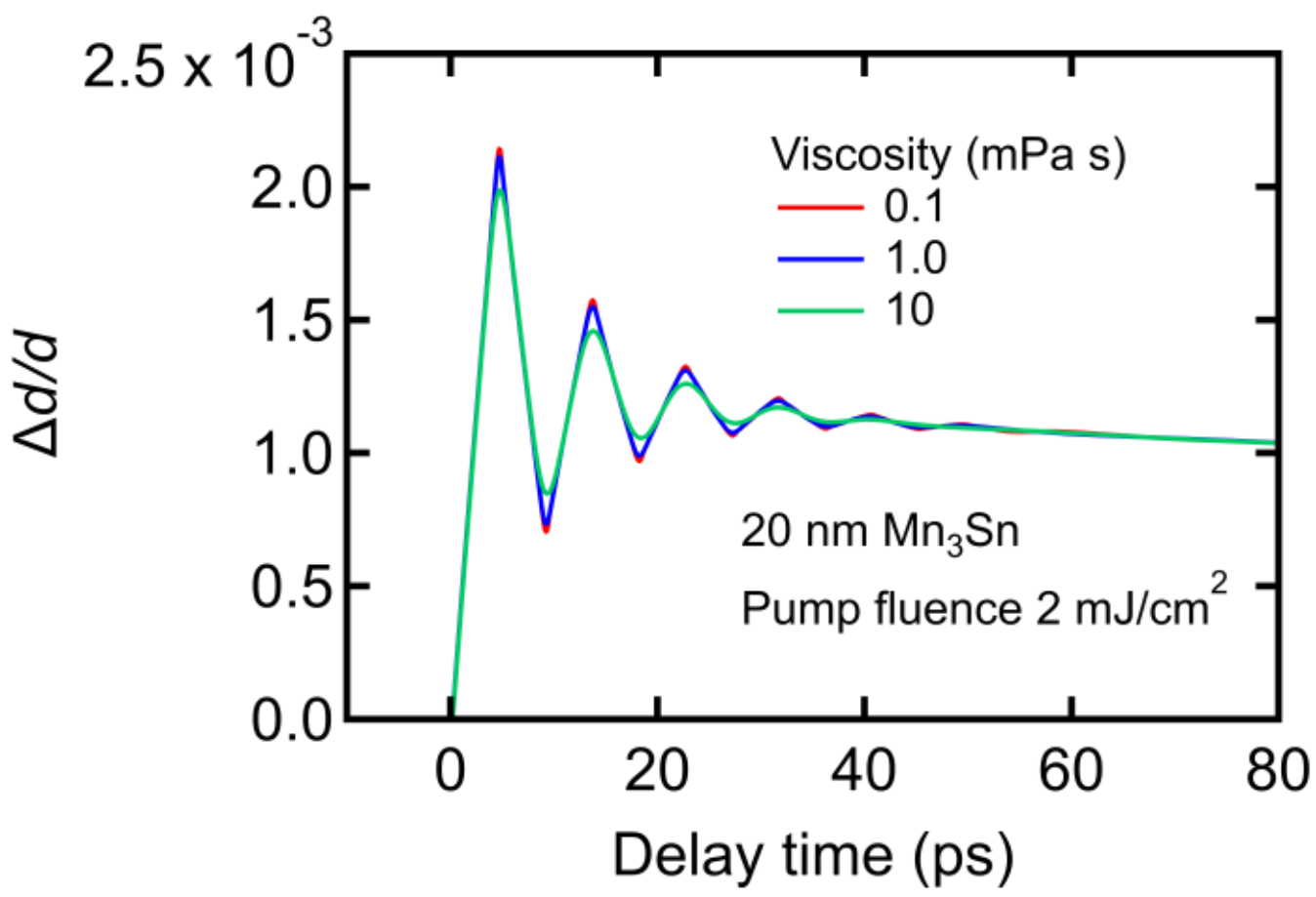

Fig. S3. $\Delta d/d$ in 20-nm sample for a pump fluence of 2 mJ cm$^{-2}$, considering different effective viscosity.

**S5. Equation for optical detection**

The theoretical model is shown by Fig. S4.

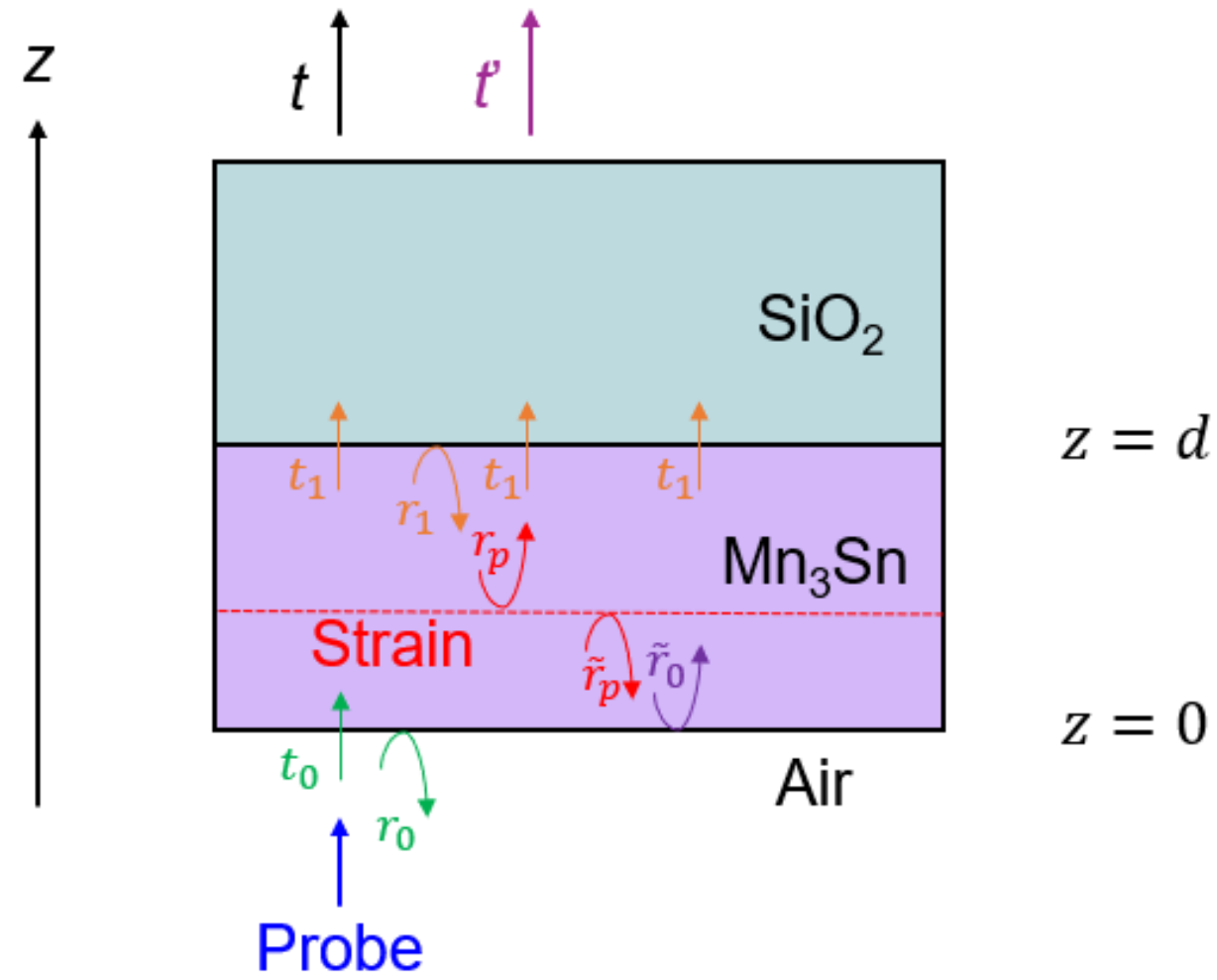

Fig. S4. Influence of strain to probe light transmittance in two-layer system.

Considering multi-reflection without perturbation, the original amplitude transmission coefficient is:

$$t = t_0 e^{i\delta_0} t_1 + t_0 r_1 \tilde{r}_0 e^{i3\delta_0} t_1 + t_0 r_1 \tilde{r}_0 r_1 \tilde{r}_0 e^{i5\delta_0} t_1 + \cdots . \quad \text{(S3)}$$

It is a geometric sequence with common ratio $q = -r_1 r_0 e^{i2\delta_0}$, thus we obtain

$$t = t_0 t_1 e^{i\delta_0} M_{MR} = \frac{t_0 t_1 e^{i\delta_0}}{1 + r_0 r_1 e^{2i\delta_0}}, \quad \text{(S4)}$$

where

$$\delta_0 = k_1 d = k_0 \tilde{n}_1 d, \quad \text{(S5)}$$

and

$$M_{MR} = \frac{1}{1 + r_0 r_1 e^{2i\delta_0}} \tag{S6}$$

is the multi-reflection factor.

With the distribution of strain induced by CAP, the amplitude transmission coefficient could be approximated by:

$$\begin{aligned} t' &= t_0 e^{i\delta} t_1 + t_0 e^{i\delta} r_1 r_p M_{MR} t_1 + t_0 \tilde{r}_p \tilde{r}_0 e^{i\delta} M_{MR} t_1 + \cdots \\ &= \frac{t_0 t_1 e^{i\delta}\left(1 + r_1 M_{MR} r_p - r_0 M_{MR} \tilde{r}_p\right)}{1 + r_0 r_1 e^{2i\delta}}, \end{aligned} \tag{S7}$$

where the first-order backscattering comes from the perturbative theory [S5-7]:

$$r_p = \frac{ik_0^2}{2k_1} \int_0^d \Delta\varepsilon(z,t) e^{2ik_1(d-z)} dz = ik_0 \frac{\partial \tilde{n}_1}{\partial \eta_{33}} \int_0^d \eta_{33}(z,t) e^{2ik_1(d-z)} dz\,, \tag{S8}$$

$$\tilde{r}_p = \frac{ik_0^2}{2k_1} \int_0^d \Delta\varepsilon(z,t) e^{2ik_1 z} dz = ik_0 \frac{\partial \tilde{n}_1}{\partial \eta_{33}} \int_0^d \eta_{33}(z,t) e^{2ik_1 z} dz, \tag{S9}$$

and the phase term:

$$\delta = \delta_0 + \Delta\delta, \tag{S10}$$

$$\Delta\delta = \frac{k_0^2}{2k_1} \int_0^d \Delta\varepsilon(z,t) dz = k_0 \frac{\partial \tilde{n}_1}{\partial \eta_{33}} \int_0^d \eta_{33}(z,t) dz. \tag{S11}$$

Omitting the second order term, we get

$$\frac{\delta t}{t} = \frac{t' - t}{t} \approx r_1 M_{MR} r_p - r_0 M_{MR} \tilde{r}_p + i\Delta\delta\left(1 - 2 r_0 r_1 e^{2i\delta_0} M_{MR}\right). \tag{S12}$$

We finally obtain

$$\begin{aligned} \frac{\Delta T}{T} = 2\mathrm{Re}\Bigg\{ ik_0 \frac{\partial \tilde{n}_1}{\partial \eta_{33}} \Bigg[ &\frac{r_1}{1 + r_0 r_1 e^{2i\delta_0}} \int_0^d \eta_{33}(z,t) e^{2ik_1(d-z)} dz \\ &- \frac{r_0}{1 + r_0 r_1 e^{2i\delta_0}} \int_0^d \eta_{33}(z,t) e^{2ik_1 z} dz + \left( \frac{1 - r_0 r_1 e^{2i\delta_0}}{1 + r_0 r_1 e^{2i\delta_0}} \right) \int_0^d \eta_{33}(z,t) dz \Bigg] \Bigg\}. \end{aligned} \tag{S13}$$

We also derive Eq. (S13) individually considering the Green function and transfer matrix.

## S6. Global fitting

Here is the detail of the global fitting process. The left-hand side of Eq. (S13) is the experimentally measured $\Delta T/T$. The right-hand side contains the photoelastic coefficient and the strain-related integral, which we can calculate by our theoretical model. For samples of different thicknesses, the phase is different, hence leading to different constants preceding the integral. We therefore treat the photoelastic coefficient as the only global fitting parameter, while all phase- and thickness-dependent terms remain individually specific.

To perform the global fitting, we simultaneously fit the 20-, 25-, 30-, and 40-nm samples and search for the solution that minimizes the mean residual across all datasets. We use an initial guess of $\partial\tilde{n}/\partial\eta_{33} = 1 + i$, and add a small random perturbation and then run the procedure repeatedly to obtain the optimal photoelastic coefficient. The fitting is performed 1000 times, and the resulting distribution is used to extract the uncertainty, as shown by the error bars in Fig. 4(b). Finally, the photoelastic coefficient was evaluated as $\partial\tilde{n}/\partial\eta_{33} \sim 23.9(\pm1.2) - i8.4(\pm1.4)$ as discussed in the main text.